\documentclass{article}

\usepackage[english]{babel}
\usepackage{biblatex} %Imports biblatex package
\usepackage{newunicodechar}

\usepackage[letterpaper,top=2cm,bottom=2cm,left=3cm,right=3cm,marginparwidth=1.75cm]{geometry}

\usepackage{amsmath}
\usepackage{graphicx}
\usepackage[colorlinks=true, allcolors=blue]{hyperref}
\usepackage{listings}
\usepackage{xcolor}
\usepackage{csquotes}
\title{Development of Data-evaluation benchmark for data Wrangling Recommendation System}
\author{Yuqing Wang, Anna Fariha}

\begin{document}
\maketitle

\begin{abstract}
In DREU summer 2024 program, I am honored
to be advised by professor Anna Fariha at the University of
Utah, in collaboration with Bhavya Chopa from Microsoft and Holden Ellsworth from the University of Utah. Our project, \href{https://afariha.github.io/papers/CoWrangler_SIGMOD_2023.pdf}{CoWrangler}\cite{Chopra}, is a data-wrangling recommender system designed to streamline data processing tasks. Recognizing that data processing is often time-consuming and complex for novice users, we aim to simplify the decision-making process regarding the most effective subsequent data operation. By analyzing over 10,000 Kaggle notebooks spanning approximately 1,000 datasets, we derive insights into common data processing strategies employed by users across various tasks. This analysis helps us understand how dataset quality influences wrangling operations, informing our ongoing efforts to possibly expand our dataset sources in the future. 
\end{abstract}

\section{Background}
\subsection{Related work}

Significant work has been done in the field that offer rich insights into data cleaning and wrangling systems, highlighting the role of algorithms in aiding data processing. Yan and He (2020)\cite{10.1145/3318464.3389738} introduced Auto-Suggest, which uses graph-like representations and min-cut clustering to recommend data preparation steps in notebooks. Bavishi et al\cite{Bavishi} discussed AutoPandas, employing neural-backed generators for program synthesis, simplifying data manipulation code generation. Guo et al. (2011)\cite{Guo} explored mixed-initiative programming in Proactive Wrangling, allowing user interactions with system suggestions for data transformation scripts. Kandel et al.(2011))\cite{Kandel} presented Wrangler, an interactive tool for visually specifying data transformations. Raza and Gulwani (2017)\cite{Raza_Gulwani_2017} focused on automated data extraction using predictive program synthesis, enhancing data preparation efficiency. These works have developed novel algorithms, particularly emphasizing the integration of graph-like recommendation systems. Each study adopts different approaches and methodologies for utilizing data; some focus solely on the input dataset, while others do not require a dataset to generate recommendations. This diversity in methods highlights a significant variance in how data dependencies are handled within the field of data wrangling.

Despite significant advancements, many current systems still struggle with efficiently providing context-aware suggestions and often fail to fully utilize the content within notebooks, focusing solely on either input or output data. Our project addresses these shortcomings by introducing an intuitive interface in combination with underlying code, enhancing user interaction and understanding. We thoroughly analyze notebook content to deliver precise recommendations and employ a theoretical framework based on the Data Quality Metric to recommend data preparation steps. The framework is informed by mixed-initiative programming models, which combine user input with automated suggestions to optimize the data wrangling process.

The goal of this project is to assist and accelerate the data wrangling process by:
\begin{enumerate}
    \item Recommending the most appropriate next transformations in real-time,
    \item Generating human-readable and efficient data-wrangling code,
    \item Enabling flexibility through human-in-the-loop interactions.
\end{enumerate}

\subsection{Technologies}
The benchmark creation utilizes the Python Abstract Syntax Tree (AST) module and the Kaggle API to enhance data manipulation capabilities. The AST module is pivotal for parsing and analyzing Python code within Jupyter notebooks. Specifically, it parses each line, tracking assignment operations and converting notebooks into a structured JSON format. This allows us to record and categorize operation types, referencing both new and existing code snippets and variables involved. Additionally, the AST module aids in extracting the variable names within method calls and navigating through the notebook to capture relevant content, which is essential for understanding how users manipulate data.

The Kaggle API plays a crucial role in accessing and downloading a vast collection of real-world data from Kaggle. It not only enables users to access datasets and associated notebooks but also provides insights into how these datasets are utilized across different notebooks. This dual-direction exploration is invaluable for obtaining a diverse range of datasets and notebooks, which are crucial for training and validating the data wrangling capabilities of our system.

\section{Method}

\subsection{Notebook conversion}

This subsection details the methodology for converting Python notebook files (.ipynb) into a structured JSON format using the Python Abstract Syntax Tree (AST) module. Our approach systematically transforms computational notebooks into a format that is amenable to further analysis and processing.

The conversion process begins by parsing the .ipynb files to extract Python code snippets. Each snippet is then represented as a JSON object structured as follows:

\begin{verbatim}
{
  "operation_type": "",
  "new_df": null,
  "prev_df": null,
  "code": "df.head(3)",
  "required_variables": {},
  "imports": {},
  "snippet_cell_number": 11
}
\end{verbatim}

The conversion process from Python notebooks to JSON format begins with the identification of import statements, which allows us to detect which modules are imported and their respective aliases or "nicknames." This initial step is crucial for ensuring that all subsequent references to these modules are correctly understood. As the parser progresses through each line, it focuses on tracking assignment operations and method calls that modify data frames, capturing both read and write operations. For each line of code that modifies the data frame, a JSON object is created to document the operation type, the new and previous states of data frames, and any variables involved in the operation. The initial output from parsing serves as input for a sequential execution, where each line is processed to observe the changes in data frames, thus providing a detailed understanding of the transformations performed by each operation. Throughout this process, dynamic updates are made to the JSON representation whenever there is an alteration in a data frame's state or variables are modified due to the data wrangling operations, ensuring an accurate tracking of the data manipulation activities within the notebooks.

\begin{figure}
    \centering
    \includegraphics[width=0.75\linewidth]{extract_variable_example.pdf}
    \caption{Example wrangling operation with variable detail}
    \label{fig:enter-label}
\end{figure}

\subsection{Web scraping}
We primarily utilize the Kaggle API to scrape notebooks and datasets. Initially, we attempted to extract notebooks using the API's search syntax to target those written in Python and to extract source data from notebook metadata. However, we encountered a rate limit after acquiring 600 notebooks. Debugging revealed that a user can access only 20 pages, with 100 notebooks per page, which was insufficient for our needs. Consequently, we pivoted to a different approach: downloading datasets first, then retrieving the notebooks that process these datasets. This method is advantageous as Kaggle hosts over 30,000 datasets, and each dataset may be associated with multiple notebooks.

 Our goal is to identify specific datasets and then download the associated notebooks for processing. To achieve this, we systematically list datasets on each page of the Kaggle API results. For each dataset listed, we download its content and metadata. If the dataset's size is below a predefined threshold—indicating a potentially higher density of data wrangling operations—we then proceed to pull its associated notebooks.

Additionally, we have implemented rigorous quality checks to refine our dataset and notebook selection process. Although notebooks with a higher number of upvotes generally signify better quality, our project specifically targets notebooks with fewer than 10 upvotes. This criterion helps us focus on notebooks that are likely to involve more extensive data wrangling tasks, which are crucial for our benchmarking needs. We also dynamically track the current pages of notebooks to avoid re-crawling the same content. Moreover, we have established a retry mechanism to manage the rate limit and connection errors effectively. This involves pausing the scraping process temporarily before restarting to prevent triggering the rate limit repeatedly and to handle potential connectivity issues smoothly.

By adopting this methodical approach, we ensure that our data collection aligns closely with our project's requirements, allowing us to gather a comprehensive and relevant set of data for analysis and benchmarking.

\subsection{Variable extraction}

After acquiring the summary output from Excel, it becomes crucial to understand the roles and meanings of variables within function calls to make sense of the project. Given the complexity of data operations, traditional code categorization methods may fall short in offering clear insights.

To address this challenge, we categorize variables into three main groups:

\begin{itemize}
    \item \textbf{Dictionaries:} Variables that function as mappings.
    \item \textbf{Functions:} Variables involved in grouping or categorizing data.
    \item \textbf{Lambda expressions:} Inline functions that often perform data manipulation on-the-fly.
\end{itemize}
For a deeper analysis, we employ the Python Abstract Syntax Tree (AST) module, which allows us to programmatically analyze and categorize code elements. Below is the Python class FunctionAnalyzer, which extends ast.NodeVisitor. This class is designed to traverse the syntax tree of Python code, identifying and categorizing functions, lambda expressions, and variables:

\begin{lstlisting}
class FunctionAnalyzer(ast.NodeVisitor):
    def __init__(self):
        self.is_function = False
        self.is_variable = False
        self.variable_value = None
        self.function_def = None
        self.result = None

    def visit_FunctionDef(self, node):
        if node.name == term:
            self.is_function = True
            self.function_def = astor.to_source(node)
            self.result = f"Function definition:\n{self.function_def}"

    def visit_Lambda(self, node):
        self.is_lambda = True
        self.lambda_def = astor.to_source(node)
        self.result = f"Lambda \hspace{1pt} function:\n{self.lambda_def}"

    def visit_Assign(self, node):
        for target in node.targets:
            if isinstance(target, ast.Name) and target.id == term:
                self.is_variable = True
                self.variable_value = astor.to_source(node.value).strip()
                self.result = f"'{term}' is a variable with value: {self.variable_value}"
        self.generic_visit(node)
\end{lstlisting}

This tool enhances our ability to dissect and understand how data is manipulated within Python notebooks, which is vital for improving the accuracy of our data-wrangling benchmarks. This approach not only clarifies the usage of variables in data operations but also supports the extraction of meaningful patterns that can guide future enhancements to the project.

\subsection{Procedures}
The benchmark creation follow a pipeline

\begin{enumerate}
\item Specify usability scores, dataset sizes, and the number of notebooks using the configuration script in our system. Then run the program to get notebook from Kaggle API
\item Utilize a Python script and convert notebook to json format 
\item Load the summarized data into an Excel sheet for further extraction and categorization of variables based on their roles in data manipulation.
\end{enumerate}

\section{Result}

A python module is developed for benchmark creation, gathering notebooks and datasets through the Kaggle API. In the initial phase, we processed 10,000 Jupyter notebooks, converting them into a structured JSON format for more detailed analysis. Our outputs include a comprehensive summary of operations, in the format of .xlsx, with specified DataFrame and dataset names, which allows for precise tracking and evaluation of data manipulations. We categorized the operations into four main types: $as\_type, \ datetime,\  apply, and\  fillna$, with special emphasis on map and apply to detail what transformations are applied or mapped onto datasets.

Variables used in these operations were classified into three types: dictionaries, functions, and lambda expressions, highlighting their effectiveness in mapping and storing data attributes. This classification reflects the diverse programming styles and objectives prevalent among Kaggle contributors.

To maintain high standards of data quality, we implemented rigorous quality control measures for the notebooks. These measures include parameters for dataset usability score, dataset size, and a requirement that the notebooks come from at least a certain number of different datasets. This approach ensures that our analysis is based on robust and representative data, facilitating more accurate and reliable benchmarks.

\section{Discussion}

The benchmark is helpful to evaluation the system and provide insight into prevalent data-wrangling practices and highlighted the diverse methodologies employed by users in their data manipulation tasks.
By categorizing operations and variables, our project not only enhances the understanding of common data-wrangling patterns but also underscores the critical role of effective data management tools in streamlining these processes. The high utilization of operations like map and apply reflects a trend towards more sophisticated, functional programming approaches in data science. Our quality control measures ensured that the benchmarks developed are both reliable and applicable to a wide array of datasets, suggesting that the recommendations made by our system are likely to be broadly relevant and beneficial. 

However, current limitations arise with the system’s handling of complex datasets, or keeping track of more than one dataset.  Though inspecting kaggle notebooks, we discovered that many are tailored for machine learning sometimes user even make dataset more comlicated for machine learning tasks. They go beyond normal use case of data cleaning as these datasets often involve intricate manipulations that exceed standard data cleaning practices. Additionally, the use of the Abstract Syntax Tree (AST) module to analyze code line by line has exposed difficulties in processing structures like for loops effectively, particularly when they include inline modifications denoted by colons.

Moving forward, enhancing the system to manage multiple datasets simultaneously and improving its ability to interpret and analyze more complex code structures will be critical.

\section{Conclusion}
In conclusion, our project has successfully developed a benchmarking system by analyzing 10,000 Python notebooks from Kaggle. Through categorizing data-wrangling operations and variable types, we have highlighted the sophisticated methods used by practitioners, underscoring the pivotal role of functional programming in data science. This understanding gives us a greater opportunity to recommend operations that precisely fit users' needs. Moving forward, tasks will include enabling our team members, who are developing the data matrix, to make sense of the gathered data and validating the system against the established benchmarks. By addressing these challenges, we aim to enhance the utility and accuracy of our recommendations, ultimately fostering more robust and efficient data science practices across various domains.

\section{Acknowledgments}
I want to express my heartfelt thanks to Professor Anna Fariha for her guidance and support throughout this project, she fostered an engaging environment for me to conquer new challenges always value my ideas. Her expertise and encouragement were crucial in navigating the complexities of data-processing practices and in the successful development of our benchmarking system. I am also thankful to my group members, Bhavya Chopra and Holden Ellsworth, for all the great teamwork and effort that went into making our project a success. Finally, I appreciate the support and the resources provided by the Computing Reseach Association, which were vital in facilitating my research as a young scholar.

\printbibliography
\end{document}